# VERITAS Observation of CTA1

Nahee Park[a] for the VERITAS Collaboration[b]

[a]University of Chicago (Enrico Fermi Institute, University of Chicago, Chicago, IL 60637, USA)
[b]http://veritas.sao.arizona.edu

**Abstract.** CTA 1 (G119.5+10.2) is a composite supernova remnant (SNR) with a shell-type structure, visible in the radio band, surrounding a smaller pulsar wind nebula. Fermi detected a radio-quiet pulsar PSR J0007+7303 within the radio shell. 26.5 hours of VERITAS observation revealed extended TeV emission from CTA 1. The centroid of the TeV emission is located near the Fermi pulsar. The integral flux of the emission was ~4% of the Crab Nebula flux (>1TeV). We present an update on the source analysis with additional exposure and possible interpretations.

## CTA1 (G119.5+10.2)

CTA1 is a composite type SNR, observed with shell type morphology in radio wavelengths and center filled morphology in X-rays. In addition to the diffuse emission extending from the center region to the radio shell, X-ray observation revealed a point source (RX J0007.0+7302) and a compact pulsar wind nebular (PWN) with a jet-like structure. From the X-ray observation, the spin down energy of the pulsar is estimated to be $\sim 4.5 \times 10^{35}$ ergs per second [1]. Deep searches in optical and radio wavelengths could not find a counterpart to RX J0007.0+7302. In the gamma-ray energy range, EGRET found a source 3EG J0010+7309 within the boundaries of the radio SNR without evidence of pulsation [2][3]. Pulsation in the GeV energy range was discovered by the Fermi Large Area Telescope, confirming RX J0007.0+7302 to be a radio quiet pulsar [4]. Recently, the Fermi LAT also reported the detection of GeV emission in the off-pulse phase interval [5]. The distance to the SNR is estimated to be $1.4 \pm 0.3$ kpc based on the association of an H1 shell on the northwest part of the remnant [6]. The age of SNR is $\sim 1.3 \times 10^4$ years [1].

The X-ray observation of CTA1 suggests that the extended non-thermal emission around the radio quiet, gamma-ray pulsar is a synchrotron PWN. Inverse Compton scattering of these synchrotron electrons with background photons can produce TeV gamma rays [7]. In fact, PWNe comprise the most populous class of TeV emitters in the galaxy [8]. Model calculations by Zhang, Jiang & Lin [9] suggest that the level of emission from the PWN in CTA1 should be detectable at TeV energies by VERITAS. The predicted gamma-ray flux in the 1 - ~30 TeV energy range was $\sim 1.1 \times 10^{-12}$ ergs cm$^{-2}$ s$^{-1}$ [9]. Observation of CTA1 by earlier imaging atmospheric Cherenkov telescopes (IACTs) provided upper limits of $2.64 \times 10^{-11}$ photons cm$^{-2}$ s$^{-1}$ (E > 250

GeV) by CAT [10], $1.25 \times 10^{-11}$ photons cm$^{-2}$ s$^{-1}$ (E > 620 GeV) by Whipple [11], and $1.09 \times 10^{-12}$ photons cm$^{-2}$ s$^{-1}$ (E > 1.3 TeV) by HEGRA [12].

## VERITAS Discovery of the Source (VER J0006+729)

VERITAS is an array of four IACTs located at the Fred Lawrence Whipple Observatory in southern Arizona [13]. Each telescope has a camera that covers a total field of view of 3.5°. The angular resolution of the array is ~0.1°, and the array is sensitive to photons with energies between 100 GeV and 30 TeV. VERITAS can detect a point source with 1 % of the flux of the Crab Nebula at the 5σ level with less than 30 hours of exposure [14].

VERITAS observed the CTA1 region for two seasons from 2010 to 2012, collecting a total of ~41 hours of exposure time. Following analysis with an integration radius of 0.23° and an energy threshold value of 1 TeV, a TeV source was found with a maximum significance of 6.5σ. The centroid of the TeV emission is located at $00^h$ $06^m$ $26^s$, +72° 59′ 1.0″, ~5 arcmin away from PSR J0007+7303. From the centroid location, the identifier of the object was designated as VER J0006+729. As can be seen in Figure 1, VER J0006+729 is extended compared to the point spread function of VERITAS. The extension can be characterized by the result of fitting a 2D asymmetric Gaussian to the data, which yields the 1σ angular extension to be 0.3° × 0.24°. The spectrum can be fit with a power-law with a spectral index of 2.2 ± 0.5. Above 1 TeV, the flux of the emission over is ~$4.0 \times 10^{-12}$ erg cm$^{-2}$ s$^{-1}$. This is ~4 % of the Crab Nebula flux above 1TeV and ~0.2 % of the pulsar spin-down luminosity if the distance is 1.4 kpc.

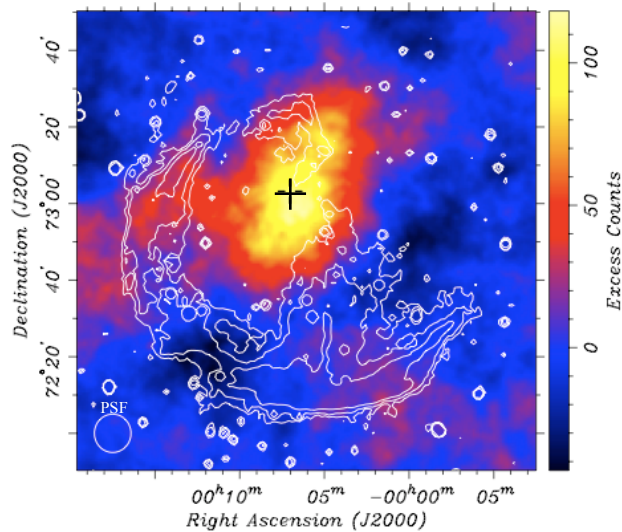

**FIGURE 1.** VERITAS excess map with circular integration area of 0.055 deg$^2$. The white contour lines show the 1420MHz radio map from the DRAO (Dominion Radio Astrophysical Observatory) synthesis telescope [6]. The cross mark indicates the position of the pulsar [4].

Good positional coincidence of VER J0006+729 to CTA1 ties the nature of the TeV source to the SNR. TeV emission from a compact type SNR can originate either from particles accelerated by the SNR shock wave, pulsar or PWN. The nature of the TeV emission in CTA1 seems to be connected to the PWN rather than the SNR shell or pulsar. Because TeV gamma-rays are good tracers for high energy electrons via inverse Compton interaction or high energy hadrons via pion decay, other TeV sources identified as connected with SNRs show either good correlation with the morphology of radio synchrotron emission or with nearby dense molecular clouds. In the case of VER J0006+729, the morphology of the TeV emission does not show any correlation with the radio shell of the SNR. There is no evidence of a dense molecular cloud found in any other wavelength including 60 μm IR or HI maps. The offset of the centroid location from the pulsar disfavors the pulsar as the origin of the emission. Also, the detection of steady emission throughout two seasons' observation suggests that the major part of the emission does not come from the pulsar. On the other hand, the well matched morphology of the non-thermal X-ray emission to the TeV emission indicates the possibility that the source of VER J0006+729 may be the PWN. Detailed modeling over a wide energy range also suggests the association to be reasonable. The properties of CTA1 agree well with the TeV/X-ray PWN population study by Kargaltsev and Pavlov [15], supporting the PWN origin of the TeV emission. At a distance of 1.4 kpc, CTA1 fits in with the picture that TeV PWNe are generally found around pulsars with ages $\lesssim 100$ kyrs and $\dot{E} \lesssim 10^{35}$ ergs s$^{-1}$.

As is apparent in Figure 1, the morphology of the TeV emission has a cometary shape, elongated toward the northwest side of the SNR where there is no strong radio emission. The asymmetric shape can be explained either by a trail of aged relativistic electrons left behind the fast moving pulsar or by the interaction of the PWN with the reverse shock of the SNR. However, the estimation of the distance that the pulsar could have moved during its lifetime is much smaller than the extent of the large scale X-ray PWN or TeV emission, even with generous assumptions about the environment variables. It is possible to explain the morphology with a scenario in which the relic PWN is being pushed to one side by the reverse shock. The overall asymmetry of the SNR radio shell suggests that the density of the interstellar medium is higher for the SE side of the remnant compared to the NW side. Thus, the reverse shock of the SNR comes likely from the SE side, making an elongation of the electron distribution toward the NW direction. This hypothesis can explain the shape of the TeV emission. A recent interaction with the reverse shock such as this may influence the properties of the thermal X-rays and their morphology. It would therefore be possible to test this hypothesis with deeper observation in X-ray wavelength.

Detailed studies and results will be published soon.

# ACKNOWLEDGMENTS

This research is supported by grants from the U.S. Department of Energy Office of Science, the U.S. National Science Foundation and the Smithsonian Institution, by NSERC in Canada, by Science Foundation Ireland (SFI 10/RFP/AST2748) and by STFC in the U.K. We acknowledge the excellent work of the technical support staff at